\documentclass{iopconfser}
\usepackage[english]{babel}
\usepackage[T1]{fontenc} 
\usepackage[utf8]{inputenc}
\usepackage{amsmath}
\usepackage{amsfonts}
\usepackage{amssymb}
\usepackage{graphicx}

\usepackage[pdftoolbar = true, pdfstartview = FitH, pdfmenubar = true]{hyperref}
\hypersetup{
    colorlinks=true,
    linkcolor=blue,
    filecolor=magenta,      
    citecolor=blue
}

\newcommand{\w}{\mathrm{w}}
\newcommand{\weff}{\mathrm{w}_{\mathrm{eff}}}

\begin{document}

\title{Friedmann cosmology with fluids and hyperfluids}

\author{Ilaria Andrei$^{a}$, Damianos Iosifidis$^{b, c}$, Laur Järv$^{a}\footnote{corresponding author}$ and Margus Saal$^{a}$}

\affil{$^a$Institute of Physics, University of Tartu, W.\ Ostwaldi 1, 50411 Tartu, Estonia \\ $^b$Scuola Superiore Meridionale, Largo San Marcellino 10, 80138 Napoli, Italy \\
$^c$ INFN Sezione di Napoli, Via Cintia, 80126 Napoli, Italy}

%\affil{}

\email{%ilaria.andrei@ut.ee, d.iosifidis@ssmeridionale.it,
laur.jarv@ut.ee%, margus.saal@ut.ee 
}

\begin{abstract}
We discuss flat Friedmann-Lema\^itre-Robertson-Walker (FLRW) metric-affine cosmology where the metric and connection as well as the matter energy-momentum and hypermomentum all obey the symmetry of spatial homogeneity and isotropy. In particular, we outline a scenario where a dark dust fluid carries spin hypermomentum which makes its effective equation of state dynamical and might relate to the DESI DR2 data.
\end{abstract}

\section{Metric-affine geometry and gravity}

In differential geometry the distance between points on a manifold is measured by a metric $g_{\mu \nu}$ while a connection $\Gamma^\lambda{}_{\mu \nu}$ defines the parallel transport of geometric objects like vectors or tensors. Although general relativity relies on the Levi-Civita connection $\tilde{\Gamma}^\lambda_{\phantom{\lambda}\mu\nu}$ derived from the metric, in principle the notions of distance (related to the metric) and direction (related to the connection) are independent of each other. Generic affine connection coefficients can be decomposed into \cite{Hehl:1994ue}
		\begin{equation}
			\label{decgamma}
			\Gamma^{\lambda}{}_{\mu \nu} = \tilde{\Gamma}^{\lambda}{}_{\mu \nu} + {\frac12 g^{\lambda\rho}\left(Q_{\mu\nu\rho} + Q_{\nu\rho\mu}
			- Q_{\rho\mu\nu}\right)} - {g^{\lambda\rho}\left(S_{\rho\mu\nu} +
			S_{\rho\nu\mu} - S_{\mu\nu\rho}\right)}\,,
		\end{equation}
where the departure from Riemannian geometry  is given in terms of nonmetricity and torsion tensors,
\begin{align}
    Q_{\alpha\mu\nu} &=-\nabla_{\alpha}g_{\mu\nu} = \partial_{\alpha}g_{\mu\nu} +          \Gamma^{\lambda}_{\phantom{\lambda}\mu\alpha}g_{\lambda\nu}
		+\Gamma^{\lambda}_{\phantom{\lambda} \nu\alpha}g_{\lambda\mu} \,, \\
    S_{\mu\nu}^{\phantom{\mu\nu}\lambda} &=\Gamma^{\lambda}_{\phantom{\lambda}[\mu\nu]} = \Gamma^{\lambda}_{\phantom{\lambda}\mu\nu} -  \Gamma^{\lambda}_{\phantom{\lambda}\nu\mu} \,,
\end{align}
with nontrivial contraction possibilities $Q_{\mu}=Q_{\mu\alpha \beta}g^{\alpha \beta}$, $\bar{Q}_{\mu}=Q_{\alpha\nu\mu}g^{\alpha\nu}$, $S_{\mu}=S_{\mu\alpha}^{\phantom{\mu\alpha} \alpha}$, and $t_{\mu}=\epsilon_{\mu\nu\alpha \beta}S^{\nu\alpha \beta}$.
In this setting it is natural to assume that matter fields can also couple to the non-Riemannian connection, thus besides the energy-momentum $T^{\alpha \beta}$ they are also characterised by hypermomentum $\Delta_{\lambda}^{\phantom{\lambda} \mu\nu}$ \cite{Hehl:1976hyperm,Hehl:1994ue},
		\begin{align}
			T^{\alpha \beta} =+\frac{2}{\sqrt{-g}}\frac{\delta(\sqrt{-g} \mathcal{L}_{M})}{\delta g_{\alpha \beta}} \,, \qquad
			\Delta_{\lambda}^{\phantom{\lambda} \mu\nu} 
			= -\frac{2}{\sqrt{-g}}\frac{\delta ( \sqrt{-g} \mathcal{L}_{M})}{\delta \Gamma^{\lambda}_{\phantom{\lambda}\mu\nu}} \,.
		\end{align}
which can be decomposed further into spin (antisymmetric), dilation (trace), and shear (symmetric traceless) parts, $\Delta_{\mu\nu\alpha}=\tau_{\mu\nu\alpha}+\frac{1}{4} g_{\mu\nu} \Delta_{\alpha} +\hat{\Delta}_{\mu\nu\alpha}$.

Perhaps the most straightforward metric-affine extension of general relativity would be given by the action
\begin{align}
			\label{S}
			S=\frac{1}{2 \kappa}\int \mathrm{d}^{4}x \sqrt{-g} \left[ R(g_{\mu\nu}, \, {{\Gamma^\lambda{}_{\mu\nu}}}) 
			+ \mathcal{L}_M \, (g_{\mu\nu}, \, {\Gamma^\lambda{}_{\mu\nu}}, \, \chi_M) \right] \,,
\end{align}
where the Ricci scalar $R$ and the Lagrangian of the matter fields $\chi_M$ depend on the general connection \eqref{decgamma}. Here we assume $\mathcal{L}_M$ is ghost free and well behaved, but leave its particular field content unspecified, as only the matter energy-momentum and hypermomentum properties exert gravitational influence. Varying the action \eqref{S} with respect to the metric and the connection gives generalised Einstein field equations and Palatini constraints,
		\begin{align}
			R_{(\mu\nu)}-\frac{1}{2}g_{\mu\nu}R &= \kappa T_{\mu\nu} \,, \label{metrf} \\
			\left( \frac{Q_{\lambda}}{2}+2 S_{\lambda}\right) g^{\mu\nu}-(Q_{\lambda}{}^{\mu\nu}
			+2 S_{\lambda}{}^{\mu\nu})+\left( \bar{Q}^{\mu}-\frac{Q^{\mu}}{2}-2 S^{\mu} \right)\delta^{\nu}_{\lambda} 
			&=\kappa \Delta_{\lambda}^{\phantom{\lambda} \mu\nu} \,. \label{conf}
		\end{align}
If the hypermomentum is zero then all non-Riemannian quantities vanish (in a certain gauge)  and \eqref{metrf} reduces to the form as in general relativity, as happens in the Palatini formulation.

\section{Homogeneous and isotropic, spatially flat cosmology}

A consistent approach to cosmology would be to assume that the metric and connection, as well as the matter energy-momentum and hypermomentum all obey cosmological symmetries, i.e.\ are homogeneous and isotropic in space. Focusing upon the spatially flat case we have \cite{Iosifidis:2020gth}
\begin{align}
			\mathrm{d}s^{2}&=-\mathrm{d}t^{2} + a^{2}(t)\delta_{ij} \mathrm{d}x^{i} \mathrm{d}x^{j} \,, \\
    Q_{\alpha \mu \nu}  &= A(t) u_\alpha h_{\mu \nu} + B(t) h_{\alpha(\mu} u_{\nu)} + C(t) u_\alpha u_\mu u_\nu \,, \quad S_{\mu\nu\alpha}=2 u_{[\mu}h_{\nu]\alpha}\Phi(t)+\epsilon_{\mu\nu\alpha\rho}u^{\rho} P(t) \,, \\
            T_{\mu\nu} &= \rho(t) \, u_{\mu}u_{\nu} + p(t) \, h_{\mu\nu} \,, \\
		\label{hypermomentum:degrees}
		\Delta_{\alpha\mu\nu} &= \phi(t) \, h_{\mu\alpha}u_{\nu} + \chi(t) \, h_{\nu\alpha}u_{\mu} + \psi(t) \, u_{\alpha}h_{\mu\nu} 
		+ \omega(t) \, u_{\alpha}u_{\mu} u_{\nu} + \epsilon_{\alpha\mu\nu\kappa}\, u^{\kappa}\zeta(t) \,,
        \end{align}
where $h_{\mu\nu} = g_{\mu\nu} + u_{\mu} u_{\nu}$ is the usual projection tensor and $u^{\mu}$ the 4-velocity of a comoving observer. The symmetry leaves 13 free functions, but some of those get related to each other by the Palatini constraint \eqref{conf} leaving only the scale factor $a$, matter density $\rho$ and pressure $p$, plus one spin component $\sigma = (\psi-\chi)/2$, and two shear components $\Sigma_{1} = (\psi+\chi)/2$, $\Sigma_{2}=(\phi+\omega)/4$ as independent quantities with nontrivial dynamics. Thus the system of cosmological equations is \cite{Andrei:2024vvy}
		\begin{subequations}
			\label{eq: FLRW equations general}
			\begin{align}
				\label{eq: FR1}
				3 H^2 &= \kappa \rho + \kappa \rho_h \,, \\
				\label{eq: FR2}
				2 \dot{H} + 3 H^2 &= - \kappa p - \kappa p_h \,, \\
				\label{eq: continuity eq}
				\dot{\rho} + 3 H (\rho + p ) &= -\dot{\rho}_h - 3 H (\rho_h + p_h ) \,,
			\end{align}
		\end{subequations}
		where $H=\dot{a}/a$ and the effective density and pressure related to the hypermomentum are
        \begin{subequations}
        \begin{align}\label{rhohph}
			\rho_h &= \frac{3 \dot{\Sigma}_2}{2} + \kappa \left(- \frac{3 \Sigma_{1} \sigma}{2} + \frac{3 \Sigma_{2}^{2}}{4} 
			- \frac{3 \Sigma_{2} \sigma}{2} - \frac{3 \sigma^{2}}{4}\right) + H \left(3 \Sigma_{1} + \frac{9 \Sigma_{2}}{2} 
			+ 3 \sigma\right)\,, \\
			p_h &= \frac{\dot{\Sigma}_2}{2} - \dot{\sigma} + \kappa \left(\Sigma_{1} \Sigma_{2} - \frac{\Sigma_{1} \sigma}{2} 
			+ \frac{3 \Sigma_{2}^{2}}{4} - \frac{\Sigma_{2} \sigma}{2} + \frac{\sigma^{2}}{4}\right) + H \left(\Sigma_{1} 
			+ \frac{3 \Sigma_{2}}{2} - 2 \sigma\right) \,. \label{pph}
		\end{align}  
        \end{subequations}

It is important to realize that this level of description is agnostic of the precise microstructure of the matter. The expressions of density, pressure, spin, and shear in terms of the fundamental fields will depend on the fundamental Lagrangian which we do not specify here. Let us only note that a generic Lagrangian formulation of hyperfluids in terms of thermodynamic quantities was developed recently in Ref.\ \cite{Iosifidis:2023kyf}. The system \eqref{eq: FLRW equations general} contains only two independent equations to determine the six dynamical variables. Hence we must impose constraints (equations of state) among the variables to solve.

For a single fluid without hypermomentum, i.e.\ in general relativity, we need one constraint, e.g.\ the constant equation of state $p = \w \rho$, to obtain for $H = \pm \sqrt{\kappa \rho/3}$ the solutions
\begin{align}
\label{eq: density GR eq}
\rho(t) &= \frac{\rho_0}{\left(1 \pm \frac{\sqrt{3\kappa \rho_0}}{2} (1+{\w_\rho}) (t-t_0) \right)^2} \,, \qquad
    H(t) = \frac{H_0}{1+\frac{3 H_0}{2}(1+{\weff})(t-t_0)} \,.
\end{align}
Here the indexes $\w_\rho=\weff=\w$ were introduced to stress the respective roles of determining how fast the energy density evolves and how fast the space expands/contracts. It turns out that a typical effect of adding hypermomentum is to make these indices to differ from each other. For instance taking a single fluid with spin hypermomentum and assuming constant equations of state $p = \w \rho$, $\sigma = b \sqrt{3 \rho/\kappa}$ results in the solutions like \eqref{eq: density GR eq}, but with $\w_\rho = \w \pm b$, $\weff = \frac{2 \w \mp b}{2 \pm 3b}$, while \eqref{eq: FR1} implies $H = \kappa \sigma/2 \pm \sqrt{\kappa \rho/3}$ \cite{Andrei:2024vvy}.

\section{Constructing dark scenarios for the Universe history}
As a complete description of cosmology needs several types of fluids, we can include in \eqref{eq: FLRW equations general} the densities $\rho_i$ and pressures $p_i$ of normal matter with $i=r, m, \Lambda$ (radiation, dust, cosmological constant), in addition to the  density $\rho_{\upsilon}$ and pressure $p_\upsilon$ of another matter component that also carries hypermomentum, 
		\begin{subequations}
			%\label{eq: FLRW equations general}
			\begin{align}
				\label{eq: FR1u}
				3 H^2 &= \kappa (\rho_r + \rho_m + \rho_\Lambda+ \rho_\upsilon) + \kappa \rho_h \,, \\
		%		\label{eq: FR2}
				2 \dot{H} + 3 H^2 &= - \kappa (p_r + p_m + p_\Lambda +p_\upsilon) - \kappa p_h \,, \\
		%		\label{eq: continuity eq r}
				\dot{\rho}_i + 3 H (\rho_i + p_i) &= 0 \,, \\
		%		\label{eq: continuity eq u}
				\dot{\rho}_{\upsilon} + 3 H (\rho_{\upsilon}+p_\upsilon)  &= -\dot{\rho}_h - 3 H (\rho_h + p_h ) \,.
			\end{align}
		\end{subequations}
The effective hypermomentum density \eqref{rhohph} and pressure \eqref{pph} define $\mathrm{w}_h=p_h/\rho_h$, while the other equations of state are with constant barotropic indexes: $p_r = \rho_r/3$, $p_m = 0$, $p_\Lambda = -\rho_\Lambda$, $p_{\upsilon} = \w_{\upsilon} \rho_{\upsilon}$.

\begin{figure}[t]
    \centering
    \includegraphics[height=7cm, trim=0cm 0cm 0cm 1.2cm, clip]{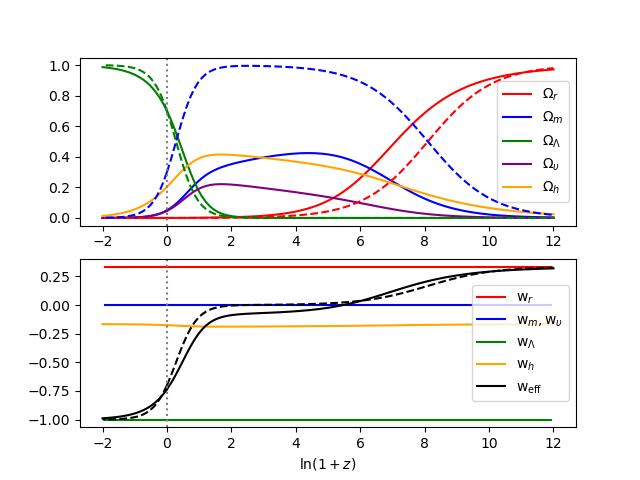}
	\caption{Cosmological evolution of the model with hypermomentum spin ($b=1/3$, $\w_\upsilon=0$).}
	\label{fig:Dust_with_spin}
\end{figure}
As a mathematically simple example, let us consider a model where the extra component is pressureless matter (dust), $\w_\upsilon = 0$, but which also carries spin in proportion to its density, $\sigma = b \sqrt{3 \rho_\upsilon/\kappa}$. For realistic behaviors and regular $p_h/\rho_h$ the dimensionless constant must better be $|b|<1$ (see \cite{Andrei:2025ykl}), we take $b=1/3$. Without specifying the respective fundamental Lagrangian here we just assume that dark matter particles have such extra property, and study the effects. We can integrate the equations numerically resulting in the plot on Fig.\ \ref{fig:Dust_with_spin} that shows the evolution of the cosmic fluids as measured in redshift $z$, depicting
\begin{align}
    \Omega_i &= \frac{\kappa \rho_i}{3 H^2} \,, \qquad \Omega_\upsilon = \frac{\kappa \rho_\upsilon}{3 H^2} \,, \qquad \Omega_h = \frac{\kappa \rho_h}{3 H^2} \,, \qquad \weff = -1 - \frac{2\dot{H}}{3H^2} \,.
\end{align}
The dashed lines represent the $\Lambda$CDM base scenario with the current relative energy densities $\Omega_{r,0}\approx 10^{-4}$, $\Omega_{\Lambda,0}=0.7$, $\Omega_{m,0}=1-\Omega_r-\Omega_\Lambda$, but $\Omega_h=\Omega_\upsilon \equiv 0$. The solid lines represent a scenario where $\Omega_{r,0}\approx 10^{-4}$, $\Omega_{\Lambda,0}=0.7$, $\Omega_{m,0}=0.05$, while the total relative energy density is still one. It means the current visible dust matter takes 5\% of the energy budget, while the remaining 25\% is accounted by the dark dust matter with spin. For the given $b$ the Friedmann equation \eqref{eq: FR1u} implies that the relative energy density of this hyperdust contributes $\Omega_{\upsilon,0}\approx0.05$ while the effective contribution arising from its spin is $\Omega_{h,0}\approx 0.2$ (changing $b$ would change these proportions). The main effect is that during the matter domination era the effective barotropic index $\weff$ is not strictly zero as in $\Lambda$CDM but rather it varies in redshift due to the interplay between the dark matter density and the contribution arising from spin. In terms of the cosmic evolution the dark matter behaves as if it had a dynamical equation of state, although the variability only comes from the spin effects. Curiously, such phenomenology may be supported by the recent DESI DR2 data \cite{DESI:2025zgx}, as several authors have recently pointed out that making the dark energy dynamical with a phantom past may be supplanted by assuming constant dark energy while making the equation of state of dark matter dynamical instead \cite{Wang:2025zri, Giani:2025hhs, Chen:2025wwn, Li:2025eqh, Li:2025dwz}. A more quantitative comparison of this model with cosmological data is the subject of our further investigation.
    
\section*{Acknowledgments}
%%%%%%%%%%%%%%%%%%%%%%%%%%%%%%%%%%%%%%%%%%%%%%%%%%%%%%%%%%%%%%%%%%%%%%%%%%%%%%%%%%%%

%This paper is based upon work from COST Action CA21136 \textit{Addressing observational tensions in cosmology with systematics and fundamental physics} (CosmoVerse) supported by COST (European Cooperation in Science and Technology). 
IA, LJ, MS were supported by the Estonian Research Council via the Center of Excellence ``Foundations of the Universe'' TK202U4 and the grant PRG2608. DI acknowledges the support of  Istituto Nazionale di Fisica Nucleare (INFN), Sezioni  di Napoli e  di Torino,  {\it Iniziative Specifiche} QGSKY.

\bibliographystyle{iopart-num.bst}
\bibliography{ref}

@article{Iosifidis:2023kyf,
    author = "Iosifidis, Damianos and Koivisto, Tomi S.",
    title = "{Hyperhydrodynamics: relativistic viscous fluids from hypermomentum}",
    eprint = "2312.06780",
    archivePrefix = "arXiv",
    primaryClass = "gr-qc",
    doi = "10.1088/1475-7516/2024/05/001",
    journal = "JCAP",
    volume = "05",
    pages = "001",
    year = "2024"
}

@article{Hehl:1976hyperm,
    author = "Hehl, F. W. and Kerlick, G. D. and Von Der Heyde, P.",
    title = "{On Hypermomentum in General Relativity. 1. The Notion of Hypermomentum}",
    journal = "Z. Naturforsch. A",
    volume = "31",
    pages = "111-114",
    year = "1976"
}

@article{Hehl:1994ue,
    author = "Hehl, Friedrich W. and McCrea, J. Dermott and Mielke, Eckehard W. and Ne'eman, Yuval",
    title = "{Metric affine gauge theory of gravity: Field equations, Noether identities, world spinors, and breaking of dilation invariance}",
    eprint = "gr-qc/9402012",
    archivePrefix = "arXiv",
    reportNumber = "TAUP-N192-94, TAUP-192-94",
    doi = "10.1016/0370-1573(94)00111-F",
    journal = "Phys. Rept.",
    volume = "258",
    pages = "1--171",
    year = "1995"
}

@article{Iosifidis:2020gth,
    author = "Iosifidis, Damianos",
    title = "{Cosmological Hyperfluids, Torsion and Non-metricity}",
    eprint = "2003.07384",
    archivePrefix = "arXiv",
    primaryClass = "gr-qc",
    doi = "10.1140/epjc/s10052-020-08634-z",
    journal = "Eur. Phys. J. C",
    volume = "80",
    number = "11",
    pages = "1042",
    year = "2020"
}

@article{Giani:2025hhs,
    author = "Giani, Leonardo and Von Marttens, Rodrigo and Piattella, Oliver Fabio",
    title = "{The matter with(in) CPL}",
    eprint = "2505.08467",
    archivePrefix = "arXiv",
    primaryClass = "astro-ph.CO",
    month = "5",
    year = "2025"
}

@article{Andrei:2024vvy,
    author = {Andrei, Ilaria and Iosifidis, Damianos and J{\"a}rv, Laur and Saal, Margus},
    title = "{Friedmann cosmology with hyperfluids}",
    eprint = "2411.19127",
    archivePrefix = "arXiv",
    primaryClass = "gr-qc",
    doi = "10.1103/PhysRevD.111.064063",
    journal = "Phys. Rev. D",
    volume = "111",
    number = "6",
    pages = "064063",
    year = "2025"
}

@inproceedings{Andrei:2025ykl,
    author = {Andrei, Ilaria and Iosifidis, Damianos and J{\"a}rv, Laur and Saal, Margus},
    title = "{Friedmann cosmology with hyperfluids of constant equation of state}",
    eprint = "2508.18479",
    archivePrefix = "arXiv",
    primaryClass = "gr-qc",
    booktitle = {Conference Proceedings for BCVSPIN 2024: Particle Physics and Cosmology in the Himalayas},
    address = {Kathmandu, Nepal, 9-13 December 2024}
}

@article{Wang:2025zri,
    author = "Wang, Deng",
    title = "{Evidence for Dynamical Dark Matter}",
    eprint = "2504.21481",
    archivePrefix = "arXiv",
    primaryClass = "astro-ph.CO",
    month = "4",
    year = "2025"
}

@article{Chen:2025wwn,
    author = "Chen, Xingang and Loeb, Abraham",
    title = "{Evolving dark energy or dark matter with an evolving equation-of-state?}",
    eprint = "2505.02645",
    archivePrefix = "arXiv",
    primaryClass = "astro-ph.CO",
    doi = "10.1088/1475-7516/2025/07/059",
    journal = "JCAP",
    volume = "07",
    pages = "059",
    year = "2025"
}

@article{Li:2025eqh,
    author = "Li, Tian-Nuo and Zhang, Yi-Min and Yao, Yan-Hong and Wu, Peng-Ju and Zhang, Jing-Fei and Zhang, Xin",
    title = "{Is non-zero equation of state of dark matter favored by DESI DR2?}",
    eprint = "2506.09819",
    archivePrefix = "arXiv",
    primaryClass = "astro-ph.CO",
    month = "6",
    year = "2025"
}

@article{Li:2025dwz,
    author = "Li, Tian-Nuo and Wu, Peng-Ju and Du, Guo-Hong and Yao, Yan-Hong and Zhang, Jing-Fei and Zhang, Xin",
    title = "{Exploring non-cold dark matter in a scenario of dynamical dark energy with DESI DR2 data}",
    eprint = "2507.07798",
    archivePrefix = "arXiv",
    primaryClass = "astro-ph.CO",
    month = "7",
    year = "2025"
}

@article{DESI:2025zgx,
    author = "Abdul Karim, M. and others",
    collaboration = "DESI",
    title = "{DESI DR2 Results II: Measurements of Baryon Acoustic Oscillations and Cosmological Constraints}",
    eprint = "2503.14738",
    archivePrefix = "arXiv",
    primaryClass = "astro-ph.CO",
    reportNumber = "FERMILAB-PUB-25-0169-PPD",
    month = "3",
    year = "2025"
}

\end{document}